\documentclass[aps,amsmath,amssymb]{revtex4}
\usepackage{graphics}
\usepackage{bm}
\usepackage{graphicx}
\usepackage{amsmath}
\usepackage{amssymb}
\usepackage{amstext}
\usepackage{amscd}
\usepackage{amsfonts}
\usepackage{appendix}
\usepackage{wasysym}
\usepackage{color}
\DeclareMathAlphabet{\EuFrak}{U}{euf}{m}{n}
\DeclareMathAlphabet{\EuScript}{U}{eus}{m}{n}

\begin{document}
\newcommand{\nd}{\noindent}
\newcommand{\nl}{\newline}
\newcommand{\be}{\begin{equation}}
\newcommand{\ee}{\end{equation}}
\newcommand{\ben}{\begin{eqnarray}}
\newcommand{\een}{\end{eqnarray}}
\newcommand{\nn}{\nonumber \\}
\newcommand{\ii}{\'{\i}}
\newcommand{\pp}{\prime}
\newcommand{\expq}{e_q}
\newcommand{\lnq}{\ln_q}
\newcommand{\quno}{q-1}
\newcommand{\qunoinv}{\frac{1}{q-1}}
\newcommand{\tr}{{\mathrm{Tr}}}

\title{{\bf  Features of constrained entropic functional variational problems}}

\author{{A. R. Plastino$^{1}$, A. Plastino$^{2,3}$, M.C.Rocca$^{2,3}$}\\
\small{$^2$Departamento de F\'{\i}sica, Fac. de Ciencias Exactas},\\
\small{Universidad Nacional de La Plata}\\
\small{C.C. 727 (1900) La Plata, Argentina}\\
\small{$^3$IFLP-CCT-CONICET-C.C 727 (1900) La Plata. Argentina}\\
\small{$^1$ CeBio y Secretaria de Investigacion,}\\
\small{Univ. Nac. del Noroeste de la Prov. de Bs. As.,}\\
\small{ UNNOBA and CONICET, R. Saenz Pe\~{n}a 456, Junin, Argentina}}



\date{\today}

\begin{abstract}

We describe in great generality features concerning  constrained entropic, functional variational problems that allow for a broad range of applications. Our discussion encompasses not only entropies but, potentially, any functional of the probability distribution, like  Fisher-information or relative entropies, etc. In particular, 
in dealing with generalized statistics in straightforward fashion one may sometimes find that 
the first thermal law $\frac{dS}{d\beta}	= \beta  \frac{d<U>}{d\beta}$ seems to be not respected. We show here that, on the contrary, it is indeed obeyed by 
any system subject to a Legendre extremization process, i.e., in all constrained entropic variational problems. 

\end{abstract}
\maketitle

\section{Introduction}

\nd Generalized entropies have become in the last 25 years a very important sub-field of statistical mechanics, with
multiple applications to many scientific disciplines \cite{BC2018,CAT2014,SANC2018,CSSS2018,BTSG2017,BSS2013,CSC2007,CAR2008,RT2007,
TB2016,GD2018,SOU2011,ARSML2015,SMCHSS2015,SMW2016,QS2013,
FR2009,LS2012}. Among the variegated set of physical scenarios to which these entropic menasures have been applied we can mention the thermostatistics of systems with long range interactions \cite{BC2018,CAT2014}, thermodynamics of many-particle systems in the overdamped motion regime \cite{SANC2018}, plasma physics \cite{CSSS2018,BTSG2017},  diverse aspects of stellar dynamics \cite{BSS2013,CSC2007}, chaotic dynamical systems \cite{CAR2008,RT2007} 
(specially, systems exhibiting weak chaos \cite{TB2016}), Bose-Einstein condensation \cite{GD2018},
thermodynamic-like description of the ground state of quantum systems \cite{SOU2011}, nonlinear Schroedinger equations \cite{ARSML2015}, speckle patterns generated by rough surfaces \cite{SMCHSS2015}, metal melting \cite{QS2013},  and the statistics of  of postural sway in humans \cite{FR2009}. Tsallis' entropy is the paramount example of a generalized entropy and the associated
thermostatistic is, by far, the one that has been most intensively investigated. The above list of recent developments
on generalized entropies and their applications (most of them concerning Tsallis entropy) is only illustrative.
In spite of the mind blowing diversity of subjects to which Tsallis theory has been applied, there actually are a few underlying basic themes that connect many of these applications. Arguably, among these common threads the three most important ones are  
(1) many-body systems with interactions whose range is of the same order as the size of the system (that is, long-range interactions), (2) systems governed by nolinear Fokker-Planck equations involving power-law diffusion terms, and (3) weak chaos. For a more detailed discussion of the vast research literature dealing with these matters see \cite{tsallis,web} and references therein. In this effort we  focus attention on the statistical derivation of  thermodynamics' first law in the guise 

\be dS/d\beta= \beta (d<U>/d\beta),\label{law}\ee
where $\beta$ is the inverse temperature, $S$ the entropy and $U$ the internal energy. This is trivial in the case of Boltzmann-Gibbs' logarithmic entropy and can be looked up in any text-book \cite{book,deslogue}. However, for general entropies such is not the case (see, as one of many possibilities, Ref. \cite{abe}). Let us look in detail at a famous example so as to clearly illustrate the problem we are talking about.

\subsection{A typical abeyance  example}

\nd An example is appropriate to appreciate the difficulties we are here referring to. 
\nd Our probability density functions (PDFs) are designed with the letter  $p$, and $p_{ME}$ would stand for the MaxEnt PDF.   

\nd We will  use the q-functions \cite{tsallis}

\be e_q(x)= [1+ (1-q)x]^{1/(1-q)};\,\,e_q(x)=\exp{(x)}\,\,for\,\,q=1 \ee  ;

\be ln_q(x)  = \frac{x^{(1-q)}-1}{1-q};\,\,ln_q(x)=\ln{(x)}\,\,for\,\,q=1. \ee
\nd We define the Tsallis q- entropy, for any real $q$, as 

\be S_T = \int dx f(p),\ee with \be f(p)=  \frac {p-p^q}{q-1}.   \ee   
Our a priori knowledge is that of  the mean energy $<U>$ (canonical ensemble). The MaxEnt variational problem   becomes, with Lagrange multipliers $\lambda_N,\,\,\lambda_U$

\be \frac {1-q p^{q-1}} {q-1}-\lambda_U U-\lambda_N=0,   \label{unounob} \ee  

\be  f'(p) = \frac {1-q p^{q-1}} {q-1}.\ee 
  \nd 
One conveniently  defines here $g(p)$ as the inverse of $ f'(p)$ such that $g[ f'(p)]=p$ \cite{universal}. 
One has 
\be g(\nu)=  q^{1-q} [1-(q-1) p']^{1/(q-1)}=  q^{1-q} e_{(2-q)}(p'). \label{xi1} \ee
It is obvious that

\be p_{ME}= g(\lambda_N + \lambda_U U), \label{xi}\ee or

\be p_{ME}= g(\lambda_N + \lambda_U U)= q^{1-q}e_{(2-q)}(\lambda_N + \lambda_U U) , \label{xi11} \ee
so that one{\it  cannot extract $\lambda_N$ from that expression. Moreover, you  do not  obtain explicitly the relation between $Z$ and $\lambda_N$}.  Since  
 one can not immediately derive from it a value for $\lambda_N$, a heuristic alternative  is to introduce

\ben  \lambda_N= -\frac{q}{q-1} Z_T^{q-1} +\frac{1}{q-1}=\frac{1}{q-1}[1-q Z_T^{q-1}], \label{rename} \een  with $Z_T$ unknown for the time being, 
and re-express $\lambda_U$ in the guise

\be  \lambda_U= q Z_T^{1-q}\beta \label{rename1}. \ee
where $\beta$ is determined by the above equation. The variational problem becomes

 \ben  \frac{1 - qp^{q-1}}{(q-1)} =  - \frac{q}{q-1} Z_T^{1-q} +\frac{1}{q-1}  +  Z_T^{1-q} q\beta U=0\\
                p^{q-1}    =  Z_T^{1-q}[1  - (q-1) \beta U],
 \een and yields 
\be  p_{ME} = 
 Z_T^{-1}  [1  - (q-1) \beta U]^{1/(q-1)},   \ee
where $\beta$ is definitely NOT the variational multiplier $\lambda_U$.  Moreover, we can now have an expression for 
\be   Z_T = \int dx\,  [1  - (q-1) \beta U]^{1/(q-1)}.      \ee
Thus, we have 
\be p^q p^{1-q} = p;\,\,p^q ln_q(p)= \frac{p - p^q}{1-q}, \ee and then 

\ben S_q=  -\int dx  p^q ln_q(p) =  \int dx p [1-p^{q-1}]/(q-1) =\\= 
  \int dx p  \left[1- (1/Z_T)^{q-1}[1  - (q-1) \beta U]\right] /(q-1) = \\ =  
	\int dx p	\left[ [1-(1/Z_T)^{q-1}]/(q-1)] +(1/Z_T)^{q-1} \beta U	\right]	 = \\ =  
	\int dx p	\left[ ln_q(Z_T) +Z_T^{1-q} \beta U 	 \right].\\ 
	S_q= ln_q(Z_T) + Z_T^{1-q} \beta \langle U \rangle=  ln_q(Z_T) + \beta  \langle U \rangle/q, \een
	entailing
	
	\be \frac{dS_q}{d\beta} = \frac{Z_T^{(2-q)/(1-q)} e_q^{2-q}(\beta U)}{1-q} + \langle U \rangle/q + 
	\beta \frac{\partial  \langle U \rangle/q }{\partial \beta}. \ee	
We encounter now, as a result, that Eq. (\ref{law}) is violated for $\beta$. This is a fact that has created some confusion in the Literature \cite{abe}.

\subsection{Our present  goal}

We will proceed, starting with the next Section, to overcome the difficulties posed by the above kind of situations. The paper is organized as follows. Section II  contains a very general proof. It applies to  any functional of the probability distribution, like generalized entropies, Fisher information, relative entropies, etc. We will demonstrate the fact that Eq. (\ref{law}) always holds, no matter what the quantifier one has in mind might be, becoming in fact a basic result of the variational problem. In order to further clarify the issue at hand we specify this proof in several 
particular instances of interest. In Section III we revisit Tsallis' quantifier. Renenyi's entropy is discussed in Sect. IV. An arbitrary, trace form entropic quantifier is the focus of Section V and, finally, an also arbitrary entropic functional lacking trace form is examined in Sect. VI. Some concluions are drawn 
in Sect. VII.

\section{The general variational problem}

\subsection{Functional derivatives: a brief reminder}
A functional $F$ of a distribution $g$ is a mapping between a collection of $g$'s and a set of numbers 
  \cite{parisi}. The functional derivative can be introduced via the Taylor expansion

\be f[g +\epsilon h]= F[g] + \epsilon \int dx\, \frac{\delta F}{\delta g(x)} h(x) + O(\epsilon^2) ,\ee
for any reasonable $h(x)$. Here $\frac{\delta F}{\delta g(x)}$ becomes the definition of a 
functional derivative.  Note that it is both a function of $x$ and a functional of $g$. In our case, generalized entropies constitute our foremost example of a functional.

\subsection{The general problem}
Let $F$ and $G$ be functionals of a normalized probability density function (PDF) $f$. 
\be F=F[f]; \,\,\, G=G[f]; \,\,\, \int dx f(x) =1. \ee

\nd Given two functionals $F$ and $G$, one wishes to extremize $F$ subject to a fixed value for $G$.  The ensuing  variational problem reads

\be \delta [F - bG - af] \,\,\,  \Rightarrow \,\,\,
 \frac{\delta F}{\delta f} - b \frac{\delta G}{\delta f}  - a =0, \ee 
while
\be   \int dx f(a,b,x) =1 \,\,\,  \Rightarrow \,\,\, \int dx [\frac{\partial f}{\partial b} +
 \frac{\partial f}{\partial a}\frac{\partial a}{\partial b}]=0. \label{mirta} \ee
Eq. (\ref{mirta}) plays a very important role in our endeavors, as we will presently see. We now face

\be \frac{dF}{db} =   \int dx \,  \frac{\delta F}{\delta f} 
[\frac{\partial f}{\partial b} +
 \frac{\partial f}{\partial a}\frac{\partial a}{\partial b}], \ee
so that, using (\ref{mirta}), as just promised
 \be \frac{dF}{db} = [b \frac{\delta G}{\delta f} + a] 
[\frac{\partial f}{\partial b} +
 \frac{\partial f}{\partial a}\frac{\partial a}{\partial b}]. \ee
Use now    $f-$normalization to derive the fundamental relation
\be \frac{dF}{db} = [b \frac{\delta G}{\delta f}] 
[\frac{\partial f}{\partial b} +
 \frac{\partial f}{\partial a}\frac{\partial a}{\partial b}]= b (dG/db) . \ee QED.
\nd The theme has been broached in different manners to ours, for example, in \cite{universal,emf,renio}, but without (i) our specific details and (2) our generality. For further clarification we address below important particular cases.

\section{Tsallis' MaxEnt variational problem revisited}
Since the Lagrange muktipliers are the focus of the problems we are trying to solve, we change notations and call them simply $a,\,b$. We have for $S_T$

\be\delta \left( \int dx \left[ \frac{f-f^{q}}{q-1} + bUf + af     \right]\right) =0, \ee
and then

\be qf^{q-1}= 1-(q-1) (a + bU), \ee
so that Tsallis' canonical  MaxEnt  distribution  $f$ with  linear constraints is

\be f= [\frac{1-[(q-1)(a + bU(x))]}{q}]^{1/(q-1)},
        \ee with $a,\,b$ Lagrange multipliers, $b$ the inverse temperature $T$.  			
				The first Law states that
				
				\be  \frac{dS}{db}	= b  \frac{d<U>}{db}.  \ee 				
				Now set
				
				\be  G= \left(1-[(q-1)(a + bU)]\right)^{(2-q)/(q-1)},  \ee
				 \be  Q(q)=    (\frac{1}{q})^{1/(q-1)}, \ee	
					\be D= \frac{1}{q-1},\ee		
							\be K= [\frac{da}{db} +  U],\ee
				entailing
				
				\be (df/db) = QD G K,    \ee  and, because of    $f-$normalization, we derive the fundamental relation
				
		\be	QD	\int dx   G K = 0. \label{cero} \ee		
		Tsallis entropy is
		
		\be S= \frac{1- \int dx f^q}{q-1},\ee so that

		\be	\frac{dS}{db}	= - Dq		\int dx  f^{q-1} D  G  K \label{uno},          \ee
		but, since
		
		\be  f^{q-1}= (1/q) \left(1-[(q-1)(a + bU)]\right).\ee
		Accordingly,
				
			\be	\frac{dS}{db}	=-QD^2 \int dx  \left(1-[(q-1)(a + bU)]\right)
			G  K, \label{one}          \ee 	
				that we decompose so as to take advantage of Eq. (\ref{cero}).
				
			\be	\frac{dS}{db}	= Q D^2 \int dx \left[
			G  K [(q-1)(a + bU)] \right] ,        \ee 	
			and re-using  Eq. (\ref{cero})

				\be	\frac{dS}{db}	=   QD \int dx  b  U G K. \label{two}      \ee
				Now:				

		\be  \frac{d<U>}{db}	=  \int dx U \frac{df}{db},  \ee
		that is
		
		\be  \frac{d<U>}{db}	= Q D\int dx U G K. \label{dos} \ee
		Comparing  Eq. (\ref{dos}) with  Eq. (\ref{two}) we see that
		
		\be  \frac{dS}{db}	= b  \frac{d<U>}{db}.  \ee
		QED.
		\section{The case of Renyi's entropy}
	 Renyi's quantifier  $S_R$ is an important quantity 
in several areas of scientific effort. One can cite as examples 
 ecology, quantum information, the Heisenberg
XY spin chain model, theoretical computer science, conformal
field theory, quantum quenching, diffusion processes, etc.  \cite{1,2,3,4,5,6,7,8,9,10},   and references therein.  An important Renyi-characteristic lies in its lack of trace form. We have	
		
\be
\label{eq2.1} S_R=\frac {1} {1-q}\ln\left(\int
f^{q}dx\right)=\frac {1} {1-q}\ln {J},\ee

\be J= \int
f^{q}dx,  \ee  with

\be \frac{dJ}{db} = q \int f^{q-1} \frac{df}{db}   dx. \label{parcialR} \ee

\nd The variational problem is

\be \delta \left(\ln\left\{\int dx  \frac{f^{q}}{(1-q)}\right\}  - \int dx\left[ bUf + af \right]\right) =0, \ee
yielding

\be \frac{qf^{q-1}}{J(1-q)} - bU  - a =0, \ee i.e.,

\be f^{q-1} = \frac{J(1-q)}{q}[a+bU],  \label{fqminusone}\ee and

\be f = \left(\frac{J(1-q)}{q}[a+bU]\right)^{1/(q-1)},\ee
is the MaxEnt solution, with $a,\,b$ Lagrange multipliers, $b$ the inverse temperature $T$.

	\be \frac{df}{db} = \frac{1}{q-1} \left(\frac{J(q-1)}{q}[a+bU]\right)^{(2-q)/(q-1)} \left(\frac{J(1-q)}{q}[\frac{da}{db} +U] + (1/q) [a+bU] (q-1) \frac{dJ}{db} 	  \right),\ee

	\be G=  \left(\frac{J(1-q)}{q}[a+bU]\right)^{(2-q)/(q-1)} ,  \ee
	
		\be K=   \left(\frac{J(1-q)}{q}[\frac{da}{db} +U] + (1/q) [a+bU] (q-1) \frac{dJ}{db} 	  \right).               \ee
	Now:
	\be \frac{df}{db}=  \frac{1}{q-1} G K ,\ee so that, on account of    $f-$normalization, we derive the fundamental relation
	
	\be \int dx G K=0.
	\label{ceroR} \ee

	\vskip 3mm 
				
				
				\nd The first Law states that
				
				\be  \frac{dS_R}{db}	= b \,  \frac{d<U>}{db},  \ee 	with

		\be  \frac{d<U>}{db}	= \int dx \, U\,  \frac{df}{db},  \ee
		that is
		
		\be  \frac{d<U>}{db}	=  \frac{1}{q-1}  \int dx\, U \, G \, K. \label{dosR} \ee	
				
				

		


\vskip 3mm \nd According to (\ref{eq2.1})

\be  \frac{dS_r}{db} = \frac{q}{J(1-q)} \int dx \,  f^{q-1} \,  \frac{df}{db},   \ee  i.e.,

\be  \frac{dS_r}{db} = \frac{q}{J(1-q)} \int dx \,  \frac{df}{db} \,  \frac{J(1-q)}{q}[a+bU]    ,   \ee or,


\be  \frac{dS_r}{db} =  \frac{1}{q-1} \int dx \,  [a+bU]  G \,K ,\ee  and using (\ref{ceroR})

\be  \frac{dS_r}{db} =  \frac{1}{q-1} \int dx \,bU\, G \,K.\ee

\nd Comparing with (\ref{dosR}) we obtain

	\be  \frac{dS_R}{db}	= b  \frac{d<U>}{db},  \ee 	QED.
				
\section{General entropies of trace form}

	\be S= \left[\int dx R[f(x)]\right],   \ee
with $R$ an arbitrary smooth function.
 Then
\be S'= \int dx R' .\ee
(Here $R'$ denotes the functional derivative). The MaxEnt variational problem is

\be \delta  \left[\int dx \left( R-af -bUf \right)  \right] =0.     \ee

\be  \left[\int dx \left(R' -a -bU \right)  \right] =0.     \ee

\be  R' = a +bU       \ee
Define now the inverse function if $R'$

\be g= (R')^{(-1)};\,\,so\,\,that\,\,\,\, g[R']=f= R'[g]. \label{inversas}  \ee

\be  f =  g[a +bU]      \ee

\be \int dy g[a +bU]=1. \ee

\be  \frac{d}{db}\int dx g[a +bU]=0.\ee
\be \int dx\, g'  [\frac{da}{db} +U] =  0.   \label{norma}\ee

\be \frac{d<U>}{db}= \int dx\,g'[a +bU]  [\frac{da}{db} +U]  U. \label{dosU} \ee
We use now (\ref{inversas}) to set 
$R'[g]=a +bU$, and then

\be \frac{dS}{db}=  \int dx\, \frac{dR}{db} =  \int dx\, R'[g] \frac{dg}{db}= 
\int dx\,(a +bU) g'  [\frac{da}{db} +U].  \ee We use now normalization (\ref{norma}) and obtain 
 the fundamental relation

\be \frac{dS}{db}=\int dx\, (a +bU) g'[a +bU]   [\frac{da}{db} +U]= \int dx\, bU g'[a +bU]   [\frac{da}{db} +U],
 \label{dosT}\ee 
so that comparing (\ref{dosU}) with (\ref{dosT}) we satisfy the first Law.

\section{General entropies lacking trace form}

		\be S= B\left(\left[\int dx R[f]\right]\right),   \ee
with $B$ an arbitrary smooth functional. Define the number $J = B'[\int dx R(f)]$.

 \be \frac{dS}{db}= J \int dx R'(df/db).\ee 
(Here $S'$ denotes the functional derivative).
Define $F=R' $ and consider the inverse function of $F$, namely,

\be g= F^{(-1)};\,\,\,F[g(f)]=f;\,\,\,g[F(f)]=f.\label{2inversas}\ee 
The MaxEnt variational problem ends up being

\be JF(f) - a -bU=0,\ee so that the MaxEnt solution's PD $f_{ME}$ is

\be  f= g[(a+ b U)/J] , \label{onebis}\ee and the MaxEnt entropy reads

\be  S_{ME}=  B\left[\int dx R[g(\{a + b U)\}/J)]\right].\label{twobis}\ee One also has

\be 0= \frac{d}{db} \int dx f .\ee 
  \be  \int dx \, g'[(a +b U)/J] \left\{ [\frac{\partial a}{\partial b}+U]J^{-1}-(1/J^2)(dJ/db)  [a+bU]
	\right\}=0,  \label{fourbis}\ee
Now,

\be \frac{d<U>}{db}= \int dx\,U g'([a +bU]/J)  \left\{ [\frac{\partial a}{\partial b}+U]J^{-1}-(1/J^2)(dJ/db)  [a+bU]
	\right\}. \label{2dosU} \ee
We will use now (\ref{2inversas}) to set 
$F^{'}[g]=(a +bU)/J$.

\be \frac {dS} {db}=J\int dx\, R^{'}[\{g(a + b U)/J\}]   g^{'}[(a +bU)/J] = 
J\int dx\,(a +bU)J^{-1} g'  \left\{ [\frac{\partial a}{\partial b}+U]J^{-1}-(1/J^2)(dJ/db)  [a+bu]
\right\}.  \ee We use now  $f-$normalization and derive the fundamental relation  
	
\be \frac{dS}{db}=J\int dx\, [(a +bU)/J] g'[(a +bU)/J]  \left\{ [\frac{\partial a}{\partial b}+U]J^{-1}-(1/J^2)(dJ/db)  [a+bU]\right\}, \ee so that

\be  \int dx\, bU g'[(a +bU)/J]  \left\{ [\frac{\partial a}{\partial b}+U]J^{-1}-(1/J^2)(dJ/db)  [a+bU]
	\right\},
 \label{2dosT}\ee 
so that comparing (\ref{2dosU}) with (\ref{2dosT}) we satisfy the first Law.

\section{Conclusions}
We have conclusively shown that the first law $\frac{dS}{d\beta}	= \beta  \frac{d<U>}{d\beta}$ is obeyed by 
any system subject to a Legendre extremization process, i.e., in any constrained entropic variational problems, no matter what form the entropy adopts and what kind of  constraints are used, 
We will demonstrate the fact that Eq. (\ref{law}) always holds, no matter what the quantifier one has in mind might be. The essential tool of our proofs is a judicious use of the normalization requirement.\vskip 3mm

\nd Note that the treatment of Section IIB encompasses the three different forms of  non-linear averaging that have been proposed for Tsallis' statistics in \cite{versions}.


\newpage

\begin{thebibliography}{99}

\bibitem{BC2018}
D. Bagchi and C. Tsallis,
Physica A {\bf 491} (2018) 869.

\bibitem{CAT2014}
L.J.L Cirto, V.R.V. Assis and C.  Tsallis,
Physica A {\bf  393} (2014) 286.

\bibitem{SANC2018} A.M.C. Souza, R.F.S. Andrade, F.D. Nobre, and E.M.F. Curado,
Physica A {\bf 491} (2018) 153.

\bibitem{CSSS2018}
B.S. Chahal, M. Singh,  Shalini and N.S. Saini,
Physica A {\bf 491} (2018) 935.

\bibitem{BTSG2017}
A.S. Bains, M. Tribeche, N.S. Saini and T.S. Gill,
Physica A {\bf 466} (2017) 111.

\bibitem{BSS2013}
E.P. Bento, J.R.P. Silva and R. Silva,
Physica A {\bf 392} (2013) 666.

\bibitem{CSC2007}
J.C. Carvalho, B.B. Soares, B.L. Canto Martins,
J.D. Do Nascimento, A. Recio-Blanco and J.R. de Medeiros,
Physica A {\bf 384} (2007) 507.

\bibitem{CAR2008}
A. Carati, Physica A {\bf 387} (2008) 1491.

\bibitem{RT2007}
G. Ruiz and C.  Tsallis, 
Physica A {\bf 386} (2007) 720.

\bibitem{TB2016}
U. Tirnakli and E.P. Borges, 
Nature Scientific Reports {\bf 6} (2016) 23644.


\bibitem{GD2018} A. Guha and P.K. Das,
Physica A {\bf 497} (2018) 272.

\bibitem{SOU2011}
A.M.C. Souza, Physica A {\bf 390} (2011) 2686.

\bibitem{ARSML2015} L.G.A. Alves, H.V. Ribeiro, M.A.F. Santos, R.S. Mendes, and E.K. Lenzi,
Physica A {\bf 429} (2015) 35.

\bibitem{SMCHSS2015}
H.C. Soares, J.B. Meireles, A.O. Castro, J.A.O. Huguenin, A.G.M. Schmidt, and L. da Silva,
Physica A {\bf 432} (2015) 1.

\bibitem{QS2013}
P. Quarati andf A.M. Scarfone,
Physica A {\bf 392} (2013) 6512.

\bibitem{FR2009}
T.D. Frank, Physica A {\bf 388} (2009) 2503.

\bibitem{LS2012} E.K. Lenzi and A.M. Scarfone, Physica A {\bf 391} (2012) 2543.

\bibitem{SMW2016}
A.M. Scarfone, H.  Matsuzoe and T. Wada, Phys. Lett. A {\ 380} (2016) 3022.







\bibitem{tsallis}  M. Gell-Mann and C. Tsallis, Eds. {\it Nonextensive Entropy:
Interdisciplinary applications}, Oxford University Press, Oxford,
2004;  C. Tsallis, {\it Introduction to Nonextensive Statistical
Mechanics: Approaching a Complex World}, Springer, New York, 2009.

\bibitem{web} See http://tsallis.cat.cbpf.br/biblio.htm for a
regularly updated bibliography on the subject.



\bibitem{book} R. B. Lindsay, H. Margenau, {\it Foundations of physics} (Dover, NY, 1957); 
F. Reif, {\it Fundamentals of Statistical and Thermal Physics} (McGraw-Hill, New York NY, 1965).

\bibitem{deslogue} R. K. Pathria, {\it Statistical Mechanics} (Pergamon, Exeter, 1993).

\bibitem{abe} S. Abe, S. Martinez, F. Pennini,
 A. Plastino, Phys.   Lett.  A {\bf 281} (2001) 126.

\bibitem{parisi} G. Parisi, {\it Statistical field theory} (Addison Wesley, New york, 1988).


\bibitem{jaynes} E.T. Jaynes, in: {\it Statistical physics}, ed. by W.K. Ford (Benjamin,
New York, 1963);
A. Katz, {\it Statistical mechanics} (Freeman, San Francisco,
1967).


\bibitem{universal} A. Plastino, A. R. Plastino, Phys. Lett. A {\bf 226} (1997) 257.






\bibitem{evaldo} E. M. F. Curado, C. Tsallis, J. Phys. A {\bf 24} (1991) L69.

\bibitem{mendes} C. Tsallis, R. S. Mendes, A.R. Plastino, Physica A {\bf 261} (1998) 534.

\bibitem{11} A. Plastino, M. C. Rocca, F. Pennini, Phys. Rev. E  {\bf 94} (2016) 012145.


\bibitem{1} C. M. Herdman, Stephen Inglis, P.-N. Roy, R. G. Melko, and A. Del
Maestro, Phys. Rev. E {\bf 90}, 013308 (2014).

\bibitem{2} Mohammad H. Ansari and Yuli V. Nazarov, Phys. Rev. B {\bf 91}, 174307
(2015).

\bibitem{3} Lei Wang and Matthias Troyer,      Phys. Rev. Lett.  {\bf 113},  110401
(2014).

\bibitem{4}  Matthew B. Hastings, Iv�n Gonz�lez, Ann B. Kallin, and Roger G. Melko, Phys. Rev. Lett {\bf 104},  157201  (2010).

\bibitem{5}  Richard Berkovits, Phys. Rev. Lett.  {\bf 115}, 206401
(2015).

\bibitem{6} Nima Lashkari,
   Phys. Rev. Lett. {\bf 113}, 051602   (2014).


\bibitem{7}  Gabor B. Halasz and Alioscia Hamma,
   Phys. Rev. Lett. {\bf 110},   170605    (2013).

\bibitem{8}  MB Hastings, I Gonz�lez, AB Kallin, RG Melko,
   Phys. Rev. Lett. {\bf 104},  157201     (2010);
   A. De Gregorio, S.M. lacus, {\bf 179}, 279 (2009).


   \bibitem{9} Leila Golshani, Einollah Pasha, Gholamhossein Yari,
   Information Sciences, {\bf 179},  2426 (2009); J.F. Bercher, Information Sciences {\bf 178},  2489 (2008).


\bibitem{10}   EK Lenzi, RS Mendes, LR da Silva,
 Physica A {\bf 280} (2000)  337.




\bibitem{skilling} J. Skilling, {\it Massive Inference and Maximum Entropy}, in Maximum Entropy and Bayesian Methods, pages 1-14, 1997 Proceedings of the 17th International Workshop on Maximum Entropy and Bayesian Methods of Statistical Analysis, Editors J. Erickson, Joshua T. Rychert,  C. Ray Smith (Springer, New York).


\bibitem{emf}  E. M. F. Curado, Braz. J. Phys.  {\bf 29} (1999) 36.

\bibitem{renio}   R.S. Mendes, Physica A  {\bf 242}  (1997) 299.


\bibitem{versions} E. M. F. Curado, C. Tsallis C,  J.
Phys. A: Math. Gen. {\bf 24} (1991) L69;  
C. Tsallis, R. S. Mendes, A. R. Plastino,  Physica A {\bf 261} (1998) 534; 
S. Martınez, F. Nicolas, F. Pennini, A. Plastino,  
Physica A {\bf 286} (2000) 489.







\end{thebibliography}
\end{document}